\documentclass{aastex}
\usepackage{emulateapj5}
\usepackage{apjfonts}

\slugcomment{UTAP-588}

\shorttitle{Distortion of UHE Sky by GMF}
\shortauthors{Takami \& Sato}

\begin{document}

\title{Distortion of Ultra-high-energy sky by Galactic Magnetic Field}

\author{Hajime Takami\altaffilmark{1} and Katsuhiko Sato\altaffilmark{1,2,3}}
\email{takami@utap.phys.s.u-tokyo.ac.jp}

\altaffiltext{1}{Department of Physics, School of Science, the University of Tokyo, 7-3-1 Hongo, Bunkyoku, Tokyo 113-0033, Japan}
\altaffiltext{2}{Research Center for the Early Universe, School of Science, the University of Tokyo, 7-3-1 Hongo, Bunkyoku, Tokyo 113-0033, Japan}
\altaffiltext{3}{Institute of Physics and Mathematics for Universe, the University of Tokyo, Kashiwa, Chiba, 277-8582, Japan}

\begin{abstract}
We investigate the deflections of UHE protons by Galactic magnetic field(GMF) 
using four conventional GMF models 
in order to discuss the positional correlation 
between the arrival distribution of UHECRs and their sources. 
UHE protons coming from the direction around the Galactic center 
are highly deflected above $8^{\circ}$ by the dipole magnetic field 
during their propagation in Galactic space. 
However, in bisymmetric spiral field models, 
there are directions with the deflection angle below $1^{\circ}$. 
One of these directions is 
toward Centaurus A, the nearest radio-loud active galactic 
nuclei that is one of possible candidates of UHECR sources. 
On the other hand, UHE protons arriving from the direction of 
the anti-Galactic center are less deflected, 
especially in bisymmetric spiral field models. 
Thus, the northern hemisphere, not including the Galactic center, 
is suitable for the studies of correlation with sources. 
The dependence on model parameters is also investigated. 
The deflection angles of UHE protons are dependent on the pitch angle 
of the spiral field. 
We also investigate distortion of the supergalactic plane by GMF. 
Since the distortion in the direction around Galactic center 
strongly depends on the GMF model, 
we can obtain information on GMF around Galactic center 
if Pierre Auger Observatory finds the significant positional correlation 
around the supergalactic plane. 
\end{abstract}


\section{Introduction} \label{intro}

The origin of ultra-high-energy cosmic rays(UHECRs) is 
one of the most intriguing problems in astroparticle physics. 
In order to reveal their nature and sources, 
UHECR observatories with larger exposures have been constructed. 
Now, the total exposure of Pierre Auger Observatory(PAO) already exceeds 
that of Akeno Giant Air Shower Array(AGASA), 
which had been the largest detector before the PAO era. 
Expectations for elucidation of UHECR sources are raised 
since PAO can detect more than 100 events per year above $4 \times 10^{19}$eV, 
which is an energy threshold that AGASA found small-scale anisotropy 
of observed UHECR arrival distribution from its 57 events.

The discussions toward UHECR astronomy have been recently began. 
There seems to be two standpoints for UHECR astronomy. 
One is study of direct correlation 
between UHECR sources and observed UHECR arrival directions. 
This can be expected 
if the Galactic magnetic field(GMF) and 
extragalactic magnetic field(EGMF) are weak 
to deflect UHECR trajectories weakly. 
However, the nature of the GMF and EGMF is 
poorly known observationally and theoretically. 
Recently, several simulations of the large-scale structure formation 
with magnetic field has described local magnetic structures 
\citep*{sigl03,dolag05}. 
\cite{sigl03} claimed that UHECR astronomy may not be possible 
if their source distribution model, magnetic structure and 
observer position are confirmed by future observation 
since UHE protons with $10^{20}$eV are even deflected 
$20^{\circ}$ or more by their structured EGMF. 
On the other hand, 
\cite{dolag05} showed that most of highest energy cosmic rays are 
very weakly deflected by EGMF 
since strong magnetic field ($\sim \mu$G) is highly localized 
at clusters of galaxies. 
Therefore, the deflection angles of UHECRs are controversial. 
The understanding of the GMF is known better than that of EGMF, but poor, 
as briefly reviewed in section \ref{gmf}. 
Number density of UHECR sources has also a crucial role. 
If the source number density is large, 
the number of observed events 
to unveil UHECR sources or source distribution is large. 
Recently, PAO reports the positional correlation 
between the arrival directions of highest energy events 
and nearby AGN\citep*{auger07}. 
AGASA also reported small-scale anisotropy of UHECR arrival distribution
within its angular resolution\citep*{takeda99}. 
These results are one of corroborating evidences for this approarch. 
The UHECR source number density is constrained 
at $\sim 10^{-5}~{\rm Mpc}^{-3}$ using small-scale anisotropy 
observed by AGASA, 
but this constraint has also large uncertainty 
due to small number of observed events
\citep*{blasi04,kachelriess05,takami07a}. 
However, this uncertainty is well reduced 
by nearly future observation \citep*{takami07a}. 
In any case, these uncertainty prevents our understanding 
of the positional correlation study at present.

The other is purely statistical approach. 
On the arrival distribution of highest energy cosmic rays 
in data combined with results of several observatories before the PAO era, 
medium-scale ($\sim 20^{\circ}$) anisotropy was found 
\citep{kachelriess06}. 
This angular scale corresponds a typical scale of event clusterings. 
The cumulative two-point auto-correlation function of the observed events 
can well reproduce that of mock data calculated 
by Monte-carlo simulation 
assuming that UHECR source distribution traces 
nearby large-scale structure within $z\simeq 0.02$ \citep*{cuoco07a}. 
They claimed that the medium-scale anisotropy reflects 
the local large-scale structure as UHECR sources. 
\cite{cuoco07b} suggested that such a statistical method 
is a powerful tool to reveal UHECR source distribution 
even if the GMF is considered. 
This medium-scale anisotropy was also found in PAO data\citep*{mollerach07}.

Our recent studies have been based on the former standpoint. 
In our previous study \citep*{takami07b}, 
we discussed the possibility that future observations 
of UHECR arrival distribution can unveil large-scale structure 
of UHECR source distribution. 
We calculated the positional correlation 
between simulated arrival distribution of UHE protons 
and their source distribution 
which reproduces the local structures, 
taking into account their propagation in a structured EGMF 
with plausible strengths. 
Our EGMF model also constructed under simple assumptions 
so that it reproduces the observed local structures. 
Taking into account uncertainties on the EGMF strength and 
the source number density, 
we investigate the positional correlation in several numbers 
of their parameters. 
Our studies using source distribution and EGMF model 
based on astronomical observations just predict the recent PAO result. 
We concluded that five year observation by PAO 
can unveil nearby UHECR source distribution at a few degree scale 
if local number density of UHECR sources is $\sim 10^{-5}~{\rm Mpc}^{-3}$. 
More observation is needed if the local number density is larger. 
The best indicator of the correlation is UHE protons above $10^{19.8}$eV.

In the previous study, 
we considered EGMF as a first step 
towards the direct positional correlation study 
between UHECR arrival directions and source positions. 
The GMF was neglected. 
However, the GMF affects the arrival directions of UHECRs 
\citep*{alvarez02,takami06,kachelriess07}. 
Thus, as a next step, 
we should discuss the deflections of UHECRs by the GMF.

In this study, 
we calculate trajectories of UHE protons in the Galactic space 
and investigate their deflections by the GMF. 
We adopt the backtracking method for calculating the trajectories. 
Trajectories of particles with the proton mass 
and the charge of -1 injected from the Earth 
can be regarded as those of extragalactic protons. 
We focus on UHE protons above $10^{19.8}$eV, 
which is best indicator of UHECR sources. 
We also investigate the distortion of extragalactic sky 
by the GMF and discuss its effect to the correlation study. 

Compositions of UHECRs is essential for the positional correlation study. 
Heavier components of UHECRs are more strongly deflected by 
magnetic fields and disturb the positional correlation. 
One of observables for the composition measurements 
is the depth of shower maximum, $X_{\rm max}$, 
which can be measured by fluorescence detectors. 
Its average value, $\left<X_{\rm max}\right>$, is sensitive to UHECR composition. 
High Resolution Fly's Eye(HiRes) reported 
that compositions of cosmic rays above $10^{19}$eV are 
dominated by protons as a result of $\left<X_{\rm max}\right>$ measurement
\citep*{abbasi05}. 
Recent result by PAO is compatible to the HiRes result 
within systematic uncertainty\citep*{unger07}. 
Another observable is muon density from extensive air shower 
of UHECR, observed by the ground array. 
Recent studies of the muon content indicate some fraction of 
heavier components in highest energy cosmic rays
\cite{engel07,glushkov07}. 
However, the interpretation of these two observables is 
dependent on hadronic interaction models, 
which include uncertainties at UHE energy region. 
On the other hand, the correlation with active galactic nuclei 
found by PAO indicates light composition. 
For the positional correlation study, proton component is a powerful tool. 
Thus, despite such uncertainties on UHECR compositions, 
we consider only protons in this study. 

This paper is organized as follows: 
in section \ref{gmf} we explain GMF models used in this study. 
In section \ref{distortion}, 
we investigate the deflection of UHE protons with $10^{19.8}$eV 
in four GMF models introduced in section \ref{gmf}. 
Dependence on parameters in the GMF models is discussed 
in section \ref{dependence}. 
In section \ref{conclusion}, 
we summarize results of the deflection 
and discuss effects of the deflection  to the correlation study.

\section{Galactic Magnetic Field} \label{gmf}

In this section, 
we briefly summarize current knowledge on the GMF 
and explain GMF models used in this study. 
For more details, see review articles \citep*{vallee04,han07}.

Faraday rotation measures of Galactic pulsars 
and extragalactic radio sources indicate 
that the GMF has two components: 
large-scale regular fields with strength of few $\mu$G 
and a turbulent or random field with comparable strength. 
The regular component includes the disk and the halo components.

The regular component in the Galactic disk 
has a pattern resembling that of the matter in the Galaxy. 
This spiral component could be well described 
by axisymmetric(ASS) or bisymmetric(BSS) field model. 
The BSS model has several number of field reversals 
and the ASS model does not. 
The halo field can be classified by parity at disk crossing. 
The odd parity:A (the even parity:S) represents 
that the spiral components in the halo fields above and below 
the Galactic plane are anti-parallel (parallel). 
From combination of the disk field models and the halo field models, 
four models are proposed. 
It has discussed which model is favored. 
\cite{vallee05} reported that a concentric ring model (like an ASS model) 
with a single field reversal and S-type parity are preferred (ASS-S model). 
On the other hand, \cite{han06} proposed that the spiral component has 
a bisymmetric structure with reversals on the boundaries of the spiral arms. 
In the halo field, 
\citep{han97,han99} suggested that A-type parity is preferable. 
This corresponds to BSS-A model. 
Both reports are based on observational results. 
However, discrimination between the models is complicated by the presence of 
smaller scale irregularities in the magnetic field, 
as well as uncertainties in the theoretical modelling. 
Thus, we adopt all four combinations 
and investigate the deflections of UHE protons in the Galactic space. 

\begin{figure*}
\epsscale{1.6}
\plotone{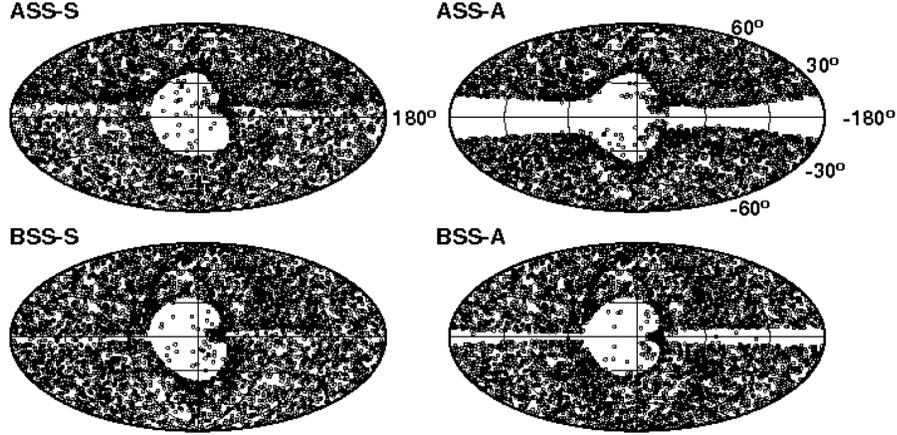}
\caption{Arrival directions of anti-protons with $10^{19.8}$eV 
injected from the Earth isotropically at 40 kpc from the Galactic center 
in the galactic coordinate. 
These can be regarded as arrival directions of extragalactic protons 
before modification by the GMF. }
\label{fig:defmap}
\end{figure*}

In the models, 
the radial and azimuthal components can be given by
\begin{equation}
B_{r_{||}} = B(r_{||},\theta) \sin p,~~~B_{\theta} = B(r_{||},\theta) \cos p. 
\end{equation}
The $r_{||}$ and $\theta$ are the Galactocentric distance 
and azimuthal angle around the Galactic center respectively 
($\theta$ is defined as increasing clockwise). 
$p$ is the pitch angle of the spiral field 
in the neighbourhood of the solar system. 
We set $p=-10^{\circ}$. 
The field strength at a point $(r_{||},\theta)$ in the Galactic plane
\citep*{stanev97} is
\begin{equation}
B(r_{||},\theta) = \left\{
\begin{array}{ll}
b(r_{||}) \cos \left( \theta - \beta \ln \frac{r_{||}}{r_0} \right) & :~{\rm BSS} \\
b(r_{||}) \left| \cos \left( \theta - \beta \ln \frac{r_{||}}{r_0} \right) \right| & :~{\rm ASS}.
\end{array}
\right. 
\end{equation}
Here, $\beta=(\tan p)^{-1}=-5.67$ and 
$r_0=10.55$ kpc is the Galactocentric distance 
of the location with maximum field strength at $l=0^{\circ}$ 
which can be expressed as 
$r_0 = (R_{\oplus} + d) \exp \left(-\frac{\pi}{2} \tan p\right)$, 
where $R_{\oplus}=8.5$ kpc is the distance of the solar system 
from the Galactic center 
and $d=-0.5$ kpc is the distance to the nearest field reversal 
from the solar system. 
Negative $d$ means that the nearest field reversal occurs 
in the direction to the Galactic center. 
$b(r_{||})$ is a radial profile of strength of magnetic field. 
The radial profile is modeled by 
\begin{equation}
b(r_{||}) = B_0 \frac{R_{\oplus}}{r_{||}}, 
\end{equation}
where $B_0=4.4\mu$G, 
which corresponds to $1.5\mu$G in the neighbourhood of the solar system. 
In the region around the Galactic center($r_{||}<4$ kpc), 
the field is highly uncertain, 
and thus assumed to be constant and equal to its value at $r_{||}=4$ kpc. 
The spiral field assumes to be zero for $r_{||}>20$ kpc.

For the spiral halo field, 
we adopt an exponentially decrease with two scale heights \citep*{stanev97} 
\begin{equation}
B(r_{||},\phi,z) = 
B(r_{||},\phi) 
\left\{
\begin{array}{ll}
\exp(-z) & :~0 \leq z \leq 0.5~{\rm kpc} \\ 
\exp(\frac{-z}{4}-\frac{3}{8}) & :~z>0.5~{\rm kpc} 
\end{array}
\right.
\label{eq:b-height}
\end{equation} 
where the factor $\exp(-3/8)$ makes the field continuous on $z$. 
The parity is represented as 
\begin{equation}
B(r_{||},\phi,-z) = \left\{
\begin{array}{ll}
B(r_{||},\phi,z) & :~{\rm S~type~parity} \\
-B(r_{||},\phi,z) & :~{\rm A~type~parity}. 
\end{array}
\right.
\end{equation}

In the GMF models with the A-type parity, 
the direction of the spiral field is reversed below the Galactic plane. 
As a result, the spiral field is discontinuous at the Galactic plane. 
It is unphysical, but one of roughly approximate models 
reflecting observational results to support the A-type parity. 
Thus, we adopt such models to investigate the deflection of UHE protons. 
Note that GMF models with A-type parity predict 
symmetric trajectories of UHE protons about the Galactic plane 
even if the dipole field, introduced just below, is included. 

Near the solar system, 
the vertical component of magnetic field with strength of $0.2-0.3\mu$G, 
directing toward the northern Galactic pole, 
is observed \citep*{han94}. 
Near the Galactic center, 
many non-thermal gaseous filaments 
perpendicular to the Galactic plane 
with tens of $\mu$G to mG have been discovered \citep*{han07}. 
These vertical magnetic fields indicate another regular component. 
In the dynamo theory, 
a dipole field is predicted with A-type parity as so-called A0 mode. 
However, in this study, 
we assume that the z-component of magnetic field is 
a dipole field as 
\begin{eqnarray}
B_x=-3~\mu_{\rm G}~{\sin\theta \cos\theta \cos\varphi}/
r^3 \nonumber \\ 
B_y=-3~\mu_{\rm G}~{\sin\theta \cos\theta \sin\varphi}/
r^3 \\
B_z=\mu_{\rm G}~{(1-3\cos^2\theta)}/r^3 ~~~~~~~ \nonumber
\label{eq:bdipole}
\end{eqnarray}
in all four models. 
Here, $\theta$ and $\varphi$ are the zenith angle and the azimuthal angle 
in the spherical coordinate centered at the Galactic center, respectively. 
$\mu_{\rm G}\sim 184.2~{\rm \mu G~kpc^3}$ is 
the magnetic moment of the Galactic dipole 
which is normalized at 0.3$\mu$G in the vicinity of the solar system.

\section{Distortion of UHE sky} \label{distortion}

In this study, we investigate the deflection of UHE protons by the GMF. 
We adopt the backtracking method for calculation 
of UHE proton trajectories in the GMF. 
We inject protons with the charge of -1 
(called {\it anti-protons} below) from the Earth isotropically, 
and follow each trajectory until the proton reaches a sphere of 
the Galactocentric radius of 40 kpc. 
The trajectory is regarded as that of a proton 
coming from extragalactic space. 
Their energy loss process is neglected 
since the energy loss length is typically much shorter 
than the propagation path length in the Galactic space. 

\begin{figure*}
\epsscale{2.1}
\plotone{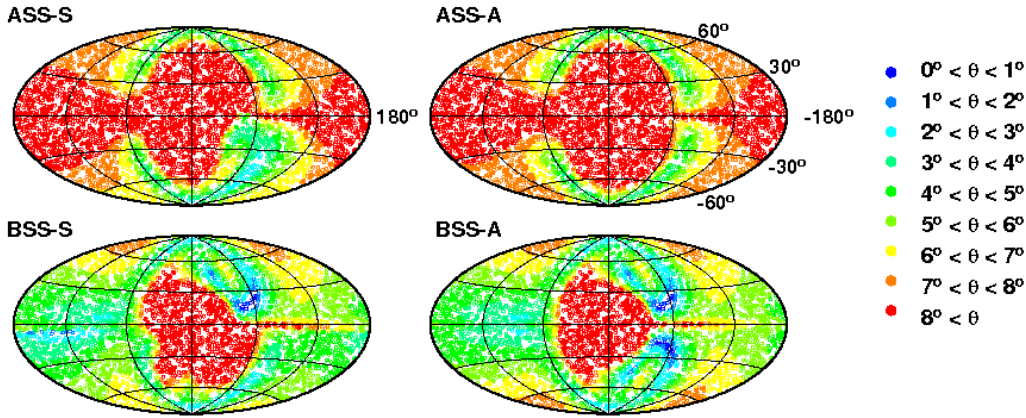}
\caption{Plots of the injection directions of anti-protons with $10^{19.8}$eV 
injected from the Earth isotropically 
with colors which represent the deflection angle 
which the anti-protons experience during their propagation in the GMF 
in the galactic coordinate. 
Each point can be regarded as the arrival direction of a proton at the Earth 
with the deflection angle during its propagation in the Galactic space. }
\label{fig:defangle}
\end{figure*}

Figure \ref{fig:defmap} shows the velocity directions of 
anti-protons with $10^{19.8}$eV injected from the Earth isotropically 
at 40 kpc from the Galactic center. 
The maps are calculated in the cases of 
the ASS-S model({\it upper-left}), 
the ASS-A model({\it upper-right}), 
the BSS-S model({\it lower-left}), 
and the BSS-A model({\it lower-right}). 
Each point is regarded as an arrival direction 
of an extragalactic proton 
before modification by the GMF, e.g. its source direction.

\begin{figure*}
\epsscale{2.0}
\plotone{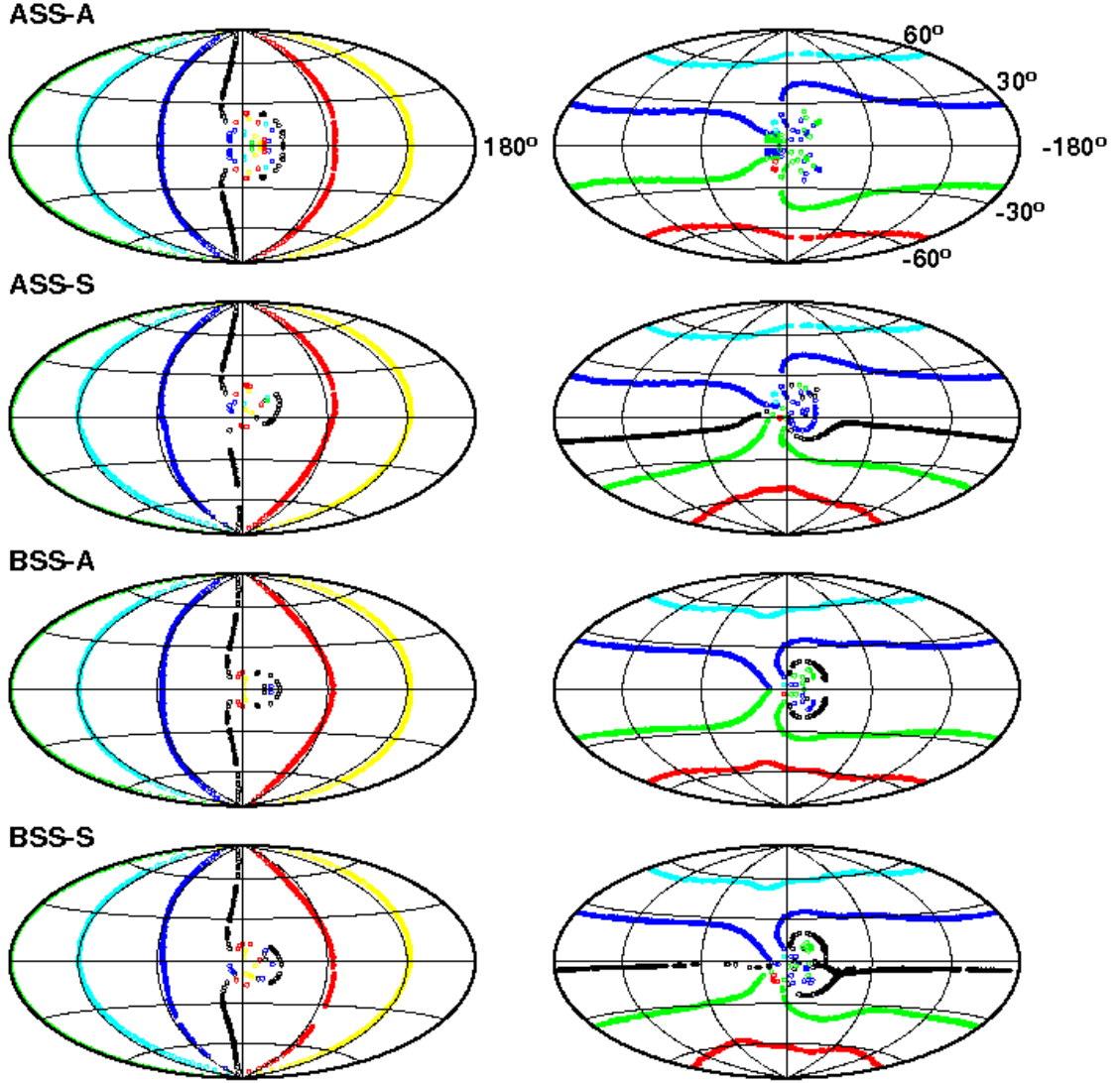}
\caption{The arrival distributions of UHE protons with $10^{19.8}$eV 
predicted from UHECR sources arranged 
along the longitude lines and the latitude lines.  
These can be regarded as the distortion of the longitude and latitude lines. 
The longitude lines considered in this figure are 
$\ell=-120^{\circ}$({\it yellow}), 
$-60^{\circ}$({\it red}), $0^{\circ}$({\it black}), 
$60^{\circ}$({\it blue}), $120^{\circ}$({\it light-blue}), 
and $180^{\circ}$({\it green}) respectively. 
The latitude lines considered in this figure are 
$b=-60^{\circ}$({\it red}), $b=-30^{\circ}$({\it green}), 
$b=0^{\circ}$({\it black}), $b=30^{\circ}$({\it blue}), and 
$b=60^{\circ}$({\it light-blue}) respectively.}
\label{fig:deflb}
\end{figure*}

All GMF models predict {\it holes}, 
where there are little arrival cosmic rays, 
with the radius of $\sim 30^{\circ}$ 
in the direction of the Galactic center. 
Anti-protons injected from the Earth are strongly scattered 
around the Galactic center due to the strong dipole field. 
Thus, protons coming from the direction of the Galactic center 
do not arrive from extragalactic sources with its direction 
(see also Fig.\ref{fig:deflb}).

In the cases of GMF models with the A-type parity, 
extragalactic protons coming from the directions around the Galactic plain 
cannot also reaches the Earth. 
The main causes are the spiral field near the solar system with A-type parity. 
The trajectories of protons arriving at the Earth 
are mainly affected by the spiral field 
in the Galactic plane and the dipole field around the Galactic center 
since strengths of these fields are relatively strong. 
Thus, anti-protons injected from the earth are, at first, deflected 
by the disk field near the solar system. 
Anti-protons injected to positive latitude are deflected 
to lower latitude in the case of injection toward the directions of 
the Galactic center, and to higher latitude 
in the case of injection toward the directions of the anti-Galactic center. 
In the both ASS models, 
the anti-protons keep on being deflected to the same directions 
due to no field reversal. 
Moreover, the A-type parity predicts a symmetric trajectories 
about the Galactic plane. 
Thus, the anti-protons are driven out from the Galactic disk region. 
On the other hand, 
the BSS models has several field reversals. 
Once the anti-protons pass over the reversal point, 
they start to be deflected to the opposite directions to those just before. 
Therefore, 
the anti-protons are driven out less strongly 
from the Galactic disk region in the BSS-A than in the ASS-A model. 
In GMF models with S-type parity, 
the apparent separation does not occur 
since the spiral field turns to the same direction 
above and below the Galactic plane. 

Next, we investigate the deflection angles of protons 
during their propagation in the Galactic space. 
In figure \ref{fig:defangle}, 
the injection directions of anti-protons with $10^{19.8}$eV are plotted 
with the deflection angles (shown as colors). 
In other words, 
a proton observed at a direction in the map has experienced deflection 
whose strength is represented as a color of the point. 
Note that the deflection angle is defined as a separation angle 
between the injection direction of an anti-proton 
and the direction of its velocity at 40 kpc from the Galactic center. 

\begin{figure*}
\epsscale{2.1}
\plotone{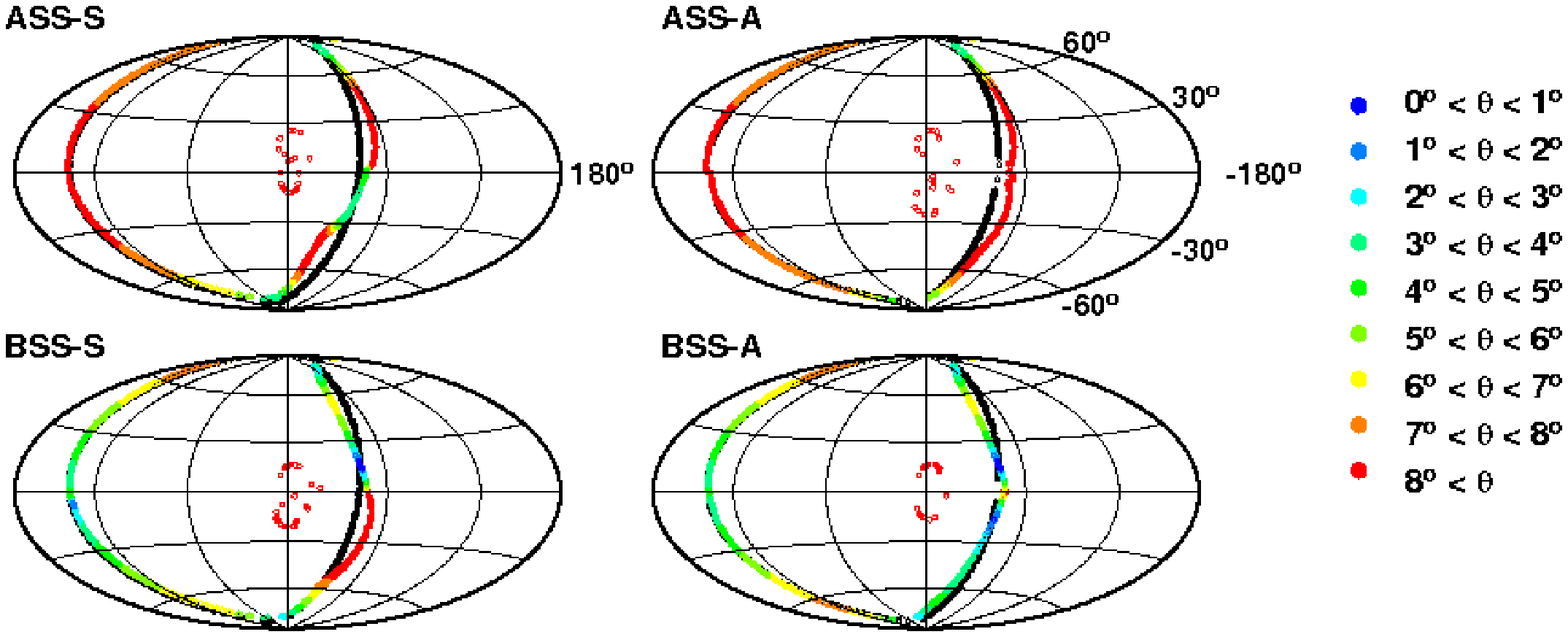}
\caption{The arrival distribution of UHE protons with $10^{19.8}$eV 
predicted from UHECR sources arranged along the supergalactic plane. 
The colors represent the deflection angles of protons 
during their propagation in the Galactic space.}
\label{fig:defsgp}
\end{figure*}

In any model, 
anti-protons injected to the direction of the Galactic center 
are strongly deflected by the dipole field near the Galactic center 
above $8^{\circ}$. 
The red regions with radius of $\sim 30^{\circ}$ exists 
even if we consider anti-protons with $10^{20.0}$eV. 
Consequently, 
protons with the arrival directions near the Galactic center 
cannot positionally correlate with their extragalactic sources 
within a few degree. 
If there is no dipole field, 
the red regions become much smaller as shown in \cite{kachelriess07}. 
We also discuss the effect of the dipole field in section \ref{dependence}. 

In the both ASS models, 
the red regions also spread around the Galactic plane 
because of reflecting no field reversal as mentioned above. 
The region with relatively small deflection angles 
reflects the direction of the spiral field near the solar system.

On the other hand, in the BSS models, 
anti-protons injected to the directions around the Galactic plane 
are less deflected than those in the ASS models, 
but the deflection angles are $3^{\circ} - 6^{\circ}$. 
If anti-protons with $10^{20.0}$eV are considered, 
the dark green regions 
(representing the deflections with $3^{\circ}-5^{\circ}$) 
change into light-blue with $1^{\circ}-3^{\circ}$. 
The deflection angles except for the directions of the Galactic center 
are approximately proportional to protons' energy. 
An intriguing structure is blue regions at around 
$(\ell,b) \simeq (-60^{\circ},20^{\circ})$ in the both BSS models, 
and $\simeq (-60^{\circ},-20^{\circ})$ in the BSS-A models. 
The protons coming from these directions are very weakly deflected 
since the deflections by the dipole field and the spiral field 
are balanced. 
In the former direction, there is Centaurus A(Cen A), 
which is the nearest radio-loud active galactic nuclei(AGNs). 
That galactic coordinate is $(l,b) \sim (-50^{\circ},20^{\circ})$. 
Radio-loud AGNs are one of strong candidates 
of UHECR sources\citep*{torres04}. 
PAO finds the positional correlation of the arrival directions 
of highest energy events with the direction of Cen A\citep*{auger07}. 
If Cen A is one of UHECR sources and those models reflect the real universe, 
spatial correlation between Cen A and the arrival directions of 
highest energy is expected. 
If the dipole field is weaker than that in this calculation, 
the blue regions are shifted to a little lower longitude 
since the dipole field deflects anti-protons injected 
to low latitude and high latitude 
to higher longitude\citep*{yoshiguchi04}.

Next, we investigate the deflection direction of extragalactic protons 
and the distortion of extragalactic sky. 
Understanding global structure of UHECR deflections is important 
on the spatial correlation study. 

Figure \ref{fig:deflb} shows the distortions of extragalactic 
longitude lines({\it left}) and latitude lines({\it right}). 
In left figures, 
we plot the injection directions of anti-protons with $10^{19.8}$eV 
whose velocity directions at 40 kpc from the Galactic center 
are $\ell=-120^{\circ}$({\it yellow}), 
$-60^{\circ}$({\it red}), $0^{\circ}$({\it black}), 
$60^{\circ}$({\it blue}), $120^{\circ}$({\it light-blue}), 
and $180^{\circ}$({\it green}) respectively. 
These lines are regarded as the arrival directions of extragalactic protons 
with $10^{19.8}$eV from sources distributed along the longitude lines. 

\begin{figure*}
\epsscale{2.1}
\plotone{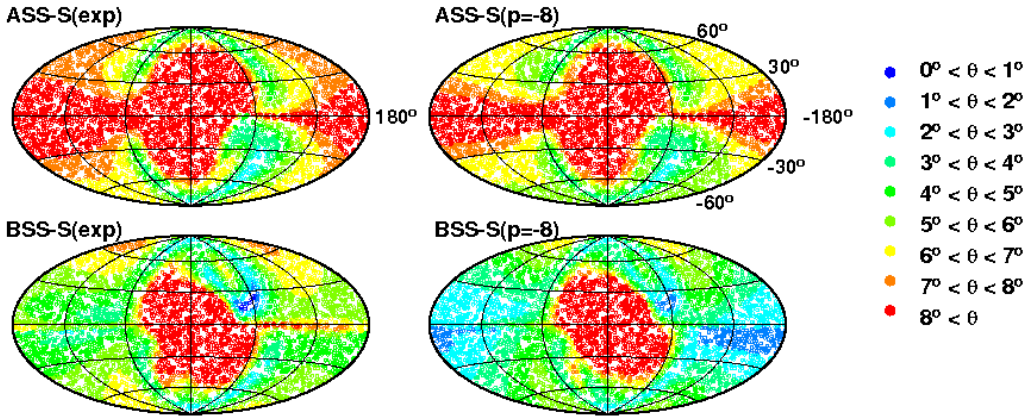}
\caption{Same as Fig.\ref{fig:defangle}, 
but in the cases of ASS-S and BSS-S models with an exponential radial profile 
of GMF strength({\it left}) and $p=-8^{\circ}$({\it right}).}
\label{fig:defmodel}
\end{figure*}

In any model, 
there are many points with different colors 
around the direction of the Galactic center. 
These are the arrival directions of protons 
which are strongly deflected by the dipole magnetic field 
around the Galactic center. 
As also shown in Fig.\ref{fig:defmap}, 
cosmic rays coming from the direction of the Galactic center 
arrive at the Earth from different directions in extragalactic space. 
These cosmic rays hardly have information on the directions of their sources. 
About 5\% of the events from the longitude lines of $60^{\circ}$ and 
$300^{\circ}$ and less than 3\% of the events from the other three lines 
arrive from the direction of the Galactic center. 
Longitude lines near the Galactic center are distorted. 
The black lines are distorted by the dipole field. 
On the other hand, 
The red lines and the blue lines are also distorted, 
but by the spiral field just inside the solar system. 
Just inside the first field reversal interior to the solar system, 
there is strong spiral field (see also Fig.1 in \cite{alvarez06}). 
The distortion of the red and blue lines depend on 
the direction of this spiral field. 
The other lines seems almost not to be distorted. 
However, protons are actually deflected along the latitude line. 
The difference of the deflection between the GMF models are 
also apparent in the right figures. 

The right figures show the distortions of five latitude lines 
with $b=-60^{\circ}$({\it red}), $b=-30^{\circ}$({\it green}), 
$b=0^{\circ}$({\it black}), $b=30^{\circ}$({\it blue}), and 
$b=60^{\circ}$({\it light-blue}) respectively. 
The same method is used to describe the plots. 
The latitude lines are highly distorted, especially around the Galactic center. 
In the ASS-A model, the black line disappears 
since protons injected from extragalactic sources along the Galactic plane 
cannot reach the Earth, as also shown in Fig.\ref{fig:defmap}. 
In the BSS-A model, 
the black line almost disappears because of the same reason 
in the ASS-A model, but there are some black points around the Galactic center. 
However, these are deflected by the dipole field 
above $10^{\circ}$ during their propagation, 
as we can see in Fig.\ref{fig:defangle}. 
Anti-protons injected from the Earth to positive latitude are 
deflected to higher latitude by the spiral field near the solar system 
in the case of the injection toward the direction of the anti-Galactic center. 
Thus, the latitude lines with positive latitude around the anti-Galactic center 
are distorted to the lower latitude. 
The deflections are stronger in the ASS models than in the BSS models 
due to no field reversal, as discussed above and see Fig.\ref{fig:defangle}. 
On the other hand, the latitude lines around the Galactic center 
are distorted to the higher latitude in the northern galactic hemisphere, 
in addition to the deflection by the dipole field. 
These tendency are found in the right figures. 
In the southern galactic hemisphere, 
the deflection directions are unchanged in GMF models with S-type parity. 
In models with A-type parity, 
symmetric pattern about the Galactic plane is predicted.

Finally, we investigate the distortion of the supergalactic plane. 
In recent years, 
PAO rejected that the hypothesis that the cosmic ray spectrum 
continues in the form of a power-law 
above an energy of $10^{19.6}$eV with 6 sigma significance 
\citep*{yamamoto07}. 
This fact, e.g. confirmation of Greisen-Zatsepin-Kuz'min(GZK) steepening
\citep*{greisen66,zatsepin66}, suggests astrophysical origins of UHECRs. 
Thus, the positional correlation 
between the arrival distribution of highest energy cosmic rays 
and the supergalactic plane is expected 
if near astrophysical objects are their sources. 
In the northern sky, 
the correlation between the arrival distribution of 
cosmic rays above $4 \times 10^{19}$eV observed 
by four surface arrays(AGASA, Yakutsk, Haverah Park, and Volcano Ranch) 
and the supergalactic plane were pointed out 
with the chance probability from the uniform distribution of 
less than 1\% \citep*{uchihori00}. 
Recent PAO result indicates such correlation\citep*{auger07}. 

We investigate the distortion of the supergalactic plane by the GMF 
in order to revisit the correlation between the arrival directions 
and the supergalactic plane. 
Figure \ref{fig:defsgp} shows the supergalactic plane({\it black points}) and 
the arrival directions of protons coming from the direction 
of the supergalactic plane({\it color plots}). 
We plot the injection directions of anti-protons 
at the Earth({\it color points}) 
whose velocity directions at 40 kpc from the Galactic center
({\it black points}) are the supergalactic plane. 
Each color shows the angle deflected by the GMF 
as shown in Fig.\ref{fig:defangle}. 

The supergalactic plane near the direction of the Galactic center 
is highly distorted by the spiral field inside the solar system 
and the dipole field around the Galactic center. 
Events from the direction of the Galactic center are 
highly deflected by the dipole field. 
In GMF models with S-type parity, 
the supergalactic plane is distorted to the opposite directions 
in the northern and southern galactic hemisphere. 
Thus, there is an intersection between the supergalactic 
plane and distorted supergalactic plane. 
The direction of the intersection has smaller deflection angle. 
In GMF models with A-type parity, 
the supergalactic plane in the southern galactic hemisphere 
is distorted like the northern galactic hemisphere. 
In the BSS-A model, 
there are two intersection with the deflection angle of below $1^{\circ}$.

The supergalactic plane near the direction of the anti-Galactic center 
is almost not distorted to the direction of longitude, 
but distorted to the direction of latitude, 
as shown by a color of each point 
and also the right figures in Fig.\ref{fig:deflb}. 
In the ASS models, 
the deflection angles are larger due to no field reversal.

We summarize our results in this section. 
Protons coming from the direction of the Galactic center 
are highly deflected by the dipole field. 
Except for the direction of the Galactic center, 
the deflection angles in the BSS models are generally smaller than 
those in the ASS models because of the field reversals. 
Extragalactic universe is distorted mainly 
along the latitude lines, 
which is reflecting local structure of the spiral field. 
In GMF models with A-type parity, 
extragalactic protons coming from the direction of the Galactic plane 
cannot reach the Earth. 
In the BSS models, 
protons coming from lower latitude are deflected smaller than higher latitude 
since the latter protons do not experience strong modifications 
of their trajectories by the field reversals. 
In the BSS models, 
there are regions that the arrival directions are almost unchanged. 
One of the regions is near Cen A, 
which is the nearest radio-loud AGN which is one of strong candidates 
of UHECR sources.

\section{Dependence on Several Parameters} \label{dependence}

In the previous section, 
we investigated the distortion of UHE sky by the GMF 
in four different GMF models with the same parameter set. 
However, the values of the parameters include some uncertainty. 
In this section, 
we discuss the dependence on such parameters.

Figure \ref{fig:defmodel} shows the same plots as Fig.\ref{fig:defangle} 
using different parameters in the ASS-S and the BSS-S model. 
A recent analysis of Faraday rotation measures suggests 
an exponential radial profile of magnetic field strength \citep*{han06}. 
According to this analysis, 
we adopt the exponential profile supposed in \cite{han06} as 
\begin{equation}
b(r) = B_0 \exp \left( \frac{-(r-R_{\oplus})}{r_b} \right), 
\end{equation}
where $r_b = 8.5$ kpc is the scale radius. 
The two left figures in Fig.\ref{fig:defmodel} represents 
the maps of the deflection angles calculated 
in the GMF models with this exponential radial profile. 
The global structure is unchanged.

The pitch angle also include some uncertainty. 
While we adopt $p=-10^{\circ}$ as the same value in previous studies 
\citep*{alvarez02,takami06}, 
the calculations by \cite{tinyakov02} and \cite{prouza03} 
adopted $p=-8^{\circ}$. 
Here, we should check the dependence on the pitch angle. 
The two right figures in Fig.\ref{fig:defmodel} shows 
the deflection angle maps calculated in the GMF models 
with the $r^{-1}$ profile with $p=-8^{\circ}$. 
$B_0$ is set to be $3.6\mu$G, which corresponds to $1.5\mu$G 
near the solar system. 

\begin{figure*}
\epsscale{1.6}
\plotone{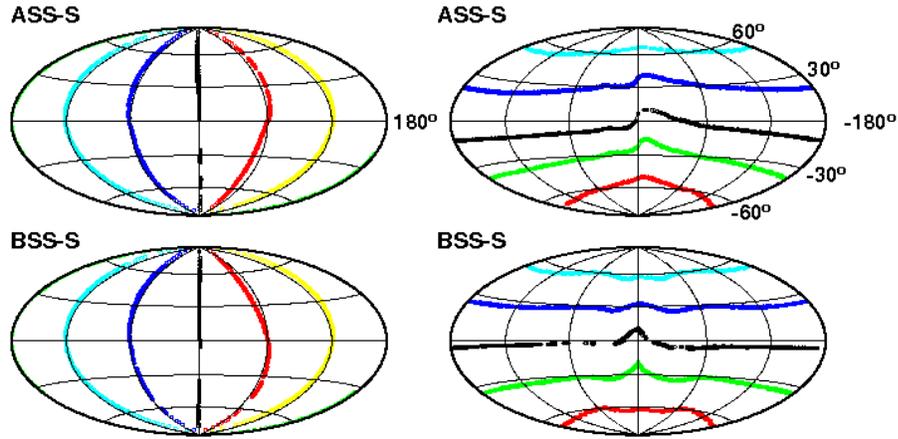}
\caption{Same as Fig.\ref{fig:deflb}, 
but in the case of no dipole field.}
\label{fig:deflb2}
\end{figure*}

An interesting change is that the deflection angles 
in the direction of the anti-Galactic center 
are smaller than those in the models with $p=-10^{\circ}$ 
in the BSS-S model. 
The reason is the distance of a field reversal 
just outside the solar system. 
In the spiral field model we use, 
the nearest field reversal interior 
to the solar system exists at $d=-0.5$ kpc, 
which is a plausible value based on many observations. 
On the other hand, the field reversals exterior to the solar system 
has weaker evidence. 
When $p$ is changed, 
the field reversal points are changed except for the nearest one. 
In particular, 
the field reversal point just outside the solar system, 
which affects the deflection angles of protons 
arriving from the direction of the anti-Galactic center, 
become nearer than that in the case of $p=-10^{\circ}$. 
Anti-protons injected to the anti-Galactic center are deflected 
to higher latitude by the spiral field in the neighbourhood of the solar system. 
When the anti-protons reaches the field reversal, 
the directions of the deflection are opposite and 
the trajectories are modified like their injection directions. 
If the field reversal is nearer, 
that modification is relatively strong. 
Thus, the deflection angles are smaller than those 
in the case of $p=-10^{\circ}$. 
In the ASS-S model, 
similar tendency can be found. 
By definition, the spiral field is weak 
near points which correspond to the field reversals in the BSS model. 
The deflection angles of anti-protons injected to the direction of 
the anti-Galactic center are smaller 
since the region with weak magnetic field nears in the case of $p=-8^{\circ}$.

Finally, we investigate the effect of the dipole field. 
Figure \ref{fig:deflb2} shows the same figure as 
Fig.\ref{fig:deflb}, 
but the dipole field is not included. 
As parameters of the spiral field, 
the same set as those in section \ref{distortion} is adopted. 
Compared to Fig.\ref{fig:deflb}, 
highly deflected events around the direction of the Galactic center disappear. 
Anti-protons are less deflected 
since the spiral field is much weaker than the dipole field 
near the Galactic center. 
However, because the magnetic strength of the spiral field 
reaches near 10 $\mu$G around the Galactic center, 
large deflections along the latitude lines are experienced 
by protons with their arrival directions of near the Galactic center. 
On the other hand, 
protons coming from the directions of the anti-Galactic center 
are almost unchanged 
since the dipole field is very weak outside the Galactic center region.

\section{Discussion \& Conclusion} \label{conclusion}

In this study, 
we investigate the deflections of UHE extragalactic protons with $10^{19.8}$eV 
in the Galactic space 
by calculating their propagation in the four different GMF models 
using the backtracking method 
in order to discuss the correlation 
between the arrival directions of UHECRs and their source positions. 

Protons arriving from the direction of the Galactic center are 
highly deflected by the dipole field. 
Source positions of such protons are quite different 
from their arrival directions. 
Thus, the Galactic center region is not suitable for the correlation study. 
If the correlation is found in the direction of the Galactic center, 
we can obtain strong constraint on the GMF around the Galactic center. 
However, in the BSS models, 
there are directions around the Galactic center 
that the arrival directions are almost unchanged. 
One of the directions is toward near Cen A, 
which is the nearest radio-loud AGN 
which is one of strong candidates of UHECR sources. 
If Cen A is confirmed as one of UHECR sources, 
the BSS models with dipole field and one of UHECR sources are confirmed.

On the other hand, 
the deflection angles of protons coming from the direction 
of the anti-Galactic center are relatively small 
for the BSS models. 
Those become smaller by adopting smaller pitch angle, $p=-8^{\circ}$ 
since the field reversal exterior to the solar system is nearer. 
Thus, the anti-Galactic center region with $|b|<30^{\circ}$ 
is best region for the correlation study. 
This region is in the northern hemisphere, 
where PAO cannot detect UHECRs. 

We also investigate the distortion of the supergalactic plane by the GMF. 
The supergalactic plane near the Galactic center is highly distorted 
and the pattern of the distortion is dependent on the spiral field model. 
\cite{ide01} showed that we can check 
whether UHECR source distribution is correlated with the supergalactic plane 
by $O(10^3)$ event detection all the sky 
This claim is based on UHECR propagation in extragalactic space. 
If the positional correlation between the arrival distribution 
and the supergalactic plane are found in the future, 
we can obtain information on the global structure of the GMF. 
On the other hand, 
the supergalactic plane near the anti-Galactic center is less distorted 
than that near the Galactic center 
although the arriving cosmic rays are deflected along the longitude line 
in the ASS models. 
In the northern hemisphere which contains the anti-Galactic center, 
the correlation between the arrival distribution and the supergalactic plane 
can be tested, which is weakly dependent on the spiral field model. 

As discussed above, 
the northern hemisphere is more suitable for the positional correlation study. 
The direction of the anti-Galactic center does not bother us 
with uncertainty of magnetic field around the Galactic center. 
In our previous study, 
we predicted the positional correlation between UHECR arrival directions 
and positions of local sources at a scale of $2^{\circ}\times2^{\circ}$ 
taking into account only EGMF\citep*{takami07b}. 
Adding the results in this study, 
we can predict the spatial correlation at a scale of $\sim 4^{\circ}$ 
for the BSS-S model with $p=-8^{\circ}$ 
and $\sim 6^{\circ}$ for the same with $p=-10^{\circ}$ 
in the direction around the anti-Galactic center. 
If the ASS model is a real situation, 
regions where we can find the positional correlation in small angular scale 
are highly restrained. 

In the northern hemisphere, 
Telescope Array is under construction 
and will start observation in the near future\citep*{fukushima07}. 
PAO also projects its northern site with larger exposure than 
its southern site\citep*{nitz07}. 
Extreme Universe Space Observatory(JEM-EUSO), 
which can detect UHECRs from International Space Station 
with extremely large exposure, is also projected\citep*{ebisuzaki07}. 
The southern site of PAO will strongly contribute to our understanding 
on UHECR nature and sources. 
As next step for UHECR astronomy, 
These detectors, which can observe in the northern hemisphere, 
will be requested. 

\acknowledgments

The work of H.T. is supported by Grants-in-Aid for JSPS Fellows. 
The work of K.S. is supported by Grants-in-Aid for 
Scientific Research provided by the Ministry of Education, 
Science and Culture of Japan through Research Grants S19104006.

\end{document}